\documentclass[fleqn,10pt]{wlscirep}
\usepackage[utf8]{inputenc}
\usepackage[T1]{fontenc}
\usepackage{lineno}
\usepackage{siunitx} 
\usepackage{hyperref}

\usepackage{multirow}       
\usepackage{bm}
\usepackage{amssymb}
\usepackage{onimage}            
\usepackage{tikz} 
\newcommand*\circled[1]{\tikz[baseline=(char.base)]{
            \node[shape=circle,draw,inner sep=2pt] (char) {#1};}}

\title{Learned denoising with simulated and experimental low-dose CT data}

\author[1,$\dag$,*]{Maximilian B. Kiss}
\author[2,$\dag$]{Ander Biguri}
\author[2]{Carola-Bibiane Schönlieb}
\author[1,3]{K. Joost Batenburg}
\author[1]{Felix Lucka}
\affil[1]{Centrum Wiskunde \& Informatica, Computational Imaging group, Amsterdam, 1098 XG, The Netherlands}
\affil[2]{University of Cambridge, DAMTP, Cambridge, CB3 0WA, United Kingdom}
\affil[3]{Leiden University, LIACS, Leiden, 2300 RA, The Netherlands}

\affil[*]{corresponding author(s): Maximilian B. Kiss (maximilian.kiss@cwi.nl)}
\affil[$\dag$]{these authors contributed equally to this work}

\begin{abstract}
Like in many other research fields, recent developments in computational imaging have focused on developing machine learning (ML) approaches to tackle its main challenges. To improve the performance of computational imaging algorithms, machine learning methods are used for image processing tasks such as noise reduction. Generally, these ML methods heavily rely on the availability of high-quality data on which they are trained. This work explores the application of ML methods, specifically convolutional neural networks (CNNs), in the context of noise reduction for computed tomography (CT) imaging. We utilize a large 2D computed tomography dataset for machine learning to carry out for the first time a comprehensive study on the differences between the observed performances of algorithms trained on simulated noisy data and on real-world experimental noisy data. The study compares the performance of two common CNN architectures, U-Net and MSD-Net, that are trained and evaluated on both simulated and experimental noisy data. The results show that while sinogram denoising performed better with simulated noisy data if evaluated in the sinogram domain, the performance did not carry over to the reconstruction domain where training on experimental noisy data shows a higher performance in denoising experimental noisy data. Training the algorithms in an end-to-end fashion from sinogram to reconstruction significantly improved model performance, emphasizing the importance of matching raw measurement data to high-quality CT reconstructions. The study furthermore suggests the need for more sophisticated noise simulation approaches to bridge the gap between simulated and real-world data in CT image denoising applications and gives insights into the challenges and opportunities in leveraging simulated data for machine learning in computational imaging.
\end{abstract}
\begin{document}

\flushbottom
\maketitle

\thispagestyle{empty}

\section*{Introduction}
\label{sec:introduction}
Computed tomography has proven itself as a powerful non-invasive imaging technique in many fields such as materials science, industrial testing, and medicine. It uses X-ray technology to create detailed cross-sectional images of the scanned object using computational methods. Since it uses harmful radiation the imposed dose on objects and patients raises concerns and safety guidelines have been established to minimize radiation exposure \cite{brenner2007computed,einstein2007estimating}. The ALARA principle \cite{mayo1997simulated}, which stands for "As Low As Reasonably Achievable" advises healthcare providers to use the lowest possible radiation dose necessary to produce high-quality images. However, the minimization of radiation dose through lowering the tube current or exposure time seriously degrades the resulting CT images if no corresponding noise compensation is applied before or during image reconstruction \cite{hsieh1998adaptive,li2004nonlinear}. Noisy images can also occur when there are for example constraints on the available time or the number of projection angles. In either setting, it is desirable to reduce the amount of noise through computational methods.
\\[6pt]
Like in many other research fields, recent developments in computational imaging have focused on developing machine learning (ML) approaches to tackle its main challenges. To improve the performance of algorithms, ML methods are used for different image processing tasks. These tasks are for example segmentation, artifact removal or noise reduction.
\\[6pt]
Generally, these ML methods heavily rely on the availability of high-quality data on which they are trained. When there is a lack of such data, usually existing data is augmented, or new data is generated artificially through simulations. These simulations mimic the problem the ML algorithms shall solve and try to resemble real-world data as good as possible.
\\[6pt]
The fundamental question arising from this approach is to which extent algorithms trained on simulated data are applicable to real-world experimental data. This work is investigating the performance of noise reduction for two common convolutional neural networks (CNNs). These networks are trained on either simulated or experimental noisy data and are applied to both experimental and simulated noisy data.
\\[6pt]
Typically, researchers would not have access to raw measurement data because CT manufacturers consider them proprietary. This severely limited both, the analysis of noise simulations but also the performance comparison of algorithms trained on simulated data \cite{whiting2006properties}. The data used in this work are 2D slices of X-ray computed tomography images published in the carefully designed study “2DeteCT – A large 2D experimental, trainable and expandable CT data collection for machine learning” \cite{Kiss_23}. This experimental data was acquired by the group for Computational Imaging at the Centrum Wiskunde \& Informatica and is openly available on zenodo \cite{kiss_maximilian_b_2023_8014758, kiss_maximilian_b_2023_8014766, kiss_maximilian_b_2023_8014787, kiss_maximilian_b_2023_8014829, kiss_maximilian_b_2023_8014874, kiss_maximilian_b_2023_8014907, kiss_maximilian_b_2023_8017583, kiss_maximilian_b_2023_8017604, kiss_maximilian_b_2023_8017612, kiss_maximilian_b_2023_8017618, kiss_maximilian_b_2023_8017624, kiss_maximilian_b_2023_8017653}. The data collection consists of 5,000 distinct image slices acquired in three different modes. The resulting images are either clean, noisy, or artifact-inflicted.
\\[6pt]
Using the paired data of clean and noisy images we create a setting for supervised learning that the CNNs can be trained on for noise reduction. Furthermore, the clean data was used as a basis for implementing the aforementioned approach of data generation through simulation, using a computationally fast, yet accurate model of the noise in an energy integrating detector. With this data collection and the newly generated data we show the difference between training on simulated noisy data and using experimental noisy data for this task.
\\[6pt]
In this paper, we utilize a large 2D computed tomography dataset for machine learning to carry out for the first time a comprehensive study on the differences between the observed performances of algorithms trained on simulated noisy data and on experimental noisy data. For this we train two common neural networks such as the generic U-Net \cite{ronneberger2015u} and the more tailored MSD-Net \cite{pelt2018mixed} on both types of noisy data, experimental and simulated. These networks are applied to the data they have been trained on but also to their respective counterparts. The evaluation follows via quantitative metrics in the sinogram and reconstructed image domain as well as qualitative visual inspection in the reconstructed image domain only.
\\[6pt]
The structure of the paper is as follows: After a brief overview of related work in noise modelling and mitigation in the field of computed tomography, we focus on the pre-processing of computed tomography data, previous noise simulation approaches and how they influenced our choices for simulating noisy training data. In the following subsections we describe the generation of our training data, the method development, the employed comparison metrics, and how we set up the computational experiments. In the results and discussion section we present the empirical selection of the noise level for our simulated noisy data and analyse the performance of the differently trained networks applied to the two types of noisy data. We focus on three aspects in our analysis: The choice for the evaluation domain, the influence of the image content, the choice of the training setting.

\section*{Methods}
\label{sec:methods}
\subsection*{Related work} \label{subsec:related}
There is a vast amount of literature investigating the theoretical derivation of accurate noise models for computed tomography images. Generally, they agree that the image noise is directly related to the imaging process and its design criteria such as exposure time, pixels size, slice width, and reconstruction algorithm \cite{faulkner1984noise}. Faulkner et al. \cite{faulkner1984noise} therefore distinguish between algorithmic and non-algorithmic contributions to noise, and between spatial as well as statistical errors in a CT scan. They note that the statistical noise in the reconstructed images is independent of the number of projections and that the uncertainty is only dependent on the total number of detected photons. Hsieh \cite{hsieh2003computed} distinguishes between two principal sources of noise in CT measurement data: quantum noise and electronic noise. Yu et al. \cite{yu2012development} showed that the latter usually can be neglected except when the number of detected photons is low and approaches the electronic noise floor. Furthermore, they emphasize that the major difficulty in simulating very-low dose CT measurement data is photon starvation artifacts. These become apparent in reconstructed image slices as ripples or rings in the central region or streaking artifacts between high-density regions. Yu et al. \cite{yu2012development} furthermore concluded that their proposed method is not able to simulate images with very-low dose because the photon starvation artifacts are quite complicated. Additionally, they reiterated the call of numerous researchers to get access to raw CT data to allow for testing algorithms for iterative reconstruction and noise reduction \cite{pan2009commercial}. Manufacturers of clinical CT scanners usually introduce nonlinear filters \cite{hsieh2003computed} on the measured data to counter beam-hardening and photon starvation artifacts. Therefore, real-world experimental raw data prior to this nonlinear filtering would enable more accurate noise simulations but are usually unavailable.
\\[6pt]
In practice, it is very challenging to bound the concept of noise in CT image reconstruction from artifacts originating from sources such as sample movement,  geometric misalignment, or under-sampling. In this work, we choose to confine our investigations purely to the noise in the sinograms induced by the photon detection in the detector. We note, however, that in reality it is hard to completely disentangle these artifacts and their origins.

\subsection*{Noise simulation}
\label{sec:sim_apps}
Zainulina et al. \cite{zainulina2022self} concluded in their work that adding noise to the images artificially could bias the predictions of a convolutional neural network (CNN) depending on the accuracy of the noise simulation which requires an in-depth understanding of the actual CT system and might not be feasible at times. The noise in low-dose CT measurement data is influenced by many factors such as the quantum noise, the logarithmic transformation of the measurements, or pre-reconstruction corrections for system calibration which makes modeling the noise in the reconstructed images particularly challenging \cite{liao2012noise}.

\paragraph{Pre-processing}
\label{sec:preprocessing}
Before CT images are reconstructed the raw sinograms that contain the measured photon counts per detector pixel usually get pre-processed using these dark-fields (D) and flat-fields (F). The sinograms (S) can then be converted into a beam intensity loss image (ILI) following the Beer-Lambert law after applying the negative logarithm to it according to the formula:
\begin{equation}
    y = - \log{(ILI)} = - \log{ \left( \frac{S - D}{F - D} \right) } 
    \label{eq:preprocessing}
\end{equation}
\noindent
Such calibrated projection data no longer follows a compound Poisson distribution but is close to a Gaussian distribution with signal-dependent variance \cite{li2004nonlinear}. Furthermore. it has been shown that particularly the logarithm operation significantly amplifies the noise when the signal is low \cite{hsieh1998adaptive}. If we want to denoise the low-dose CT measurements before reconstruction this is best done in the stage of the beam intensity loss image (ILI), so before taking the negative logarithm. If we would denoise the X-ray absorption sinogram ($y$) instead of the beam intensity loss image (ILI), the application of the negative logarithm would have amplified the noise and changed its distribution. 
\\[6pt]
Based on these findings and considerations we pre-process the raw sinograms of the experimental noisy measurements to beam intensity loss images (ILI) as shown in Eq.~\ref{eq:preprocessing} and apply the denoising before taking the negative logarithm. For the generation of the simulated noisy sinograms we pre-process the sinograms of the high-dose CT measurements of "mode 2" with dark- and flat-field corrections before applying simulated noise to them. With this we prevent our data to experience distributional shifts that might influence the performance of the denoising networks.

\paragraph{Noise simulation approaches}
To date, there have been proposed several different approaches to simulate the noise in CT measurement data \cite{frush2002computer,mayo1997simulated,amir2003dose,benson2010synthetic,whiting1998image,whiting2002signal,whiting2006properties,lasio2007statistical,massoumzadeh2005noise,massoumzadeh2009validation}. Under the condition that the raw data of a high-dose and low-noise scan is available many studies simulated low-dose projection data by applying synthetic Poisson noise or a combination of synthetic Poisson and Gaussian noise to a high-dose scan \cite{elbakri2002statistical, elbakri2003efficient}. The common models for noise simulation use a relatively simple model of CT acquisition considering a monochromatic X-ray source that generates photons that are attenuated by a scanned object and detectors counting surviving photons which are governed by Poisson statistics. More complicated methods range from a detailed characterization of signal statistics of X-ray CT \cite{whiting1998image,whiting2002signal,whiting2006properties} over noise equivalent quanta \cite{massoumzadeh2005noise,massoumzadeh2009validation} to accounting for energy-integrating detectors \cite{lasio2007statistical, massoumzadeh2009validation}. The interested reader may be pointed to the study of Zabic et al. giving a broad overview on the state-of-the-art \cite{vzabic2013low}. 
\\[6pt]
To motivate our noise simulation approach we highlight what approaches have been used in practice by previous publications in the field. In particular, there are three noise challenges that have been conducted in the past ten years that have attracted attention to deep learning based denoising. Firstly, the Mayo clinic low-dose CT challenges of 2016 \cite{mccollough2016tu} and of 2021 \cite{Mayo_2021} which encompass 30 and 300 patient scans respectively of roughly 70 slices each with noisy reconstruction and projection data simulated from clean reconstructed volumes. Secondly, the LoDoPaB-CT dataset \cite{leuschner2021lodopab} which uses 800 patient scans selected from the LIDC/IDRI database and contains over 40,000 scan slices. Thirdly, the IEEE ICASSP Grand Challenge 8 \cite{ICASSP_SPGC8} which also utilizes the LIDC/IDRI database and contains 1010 3D cone-beam CT (CBCT) images. Whereas the first two publications simulate their noisy data only by applying Poisson noise to the projection data, the third generates CBCT projection data with a custom noise simulator that accounts for photon counts, flat-fields, electronic sources, and detector cross-talk as sources of noise. Similar approaches have been undertaken by Bruno de Man et al. from GE research \cite{de2007catsim, wu2022xcist} and Jingyan Xu and Benjamin M. W. Tsui \cite{xu2009electronic} and shall be the basis for this work's noise simulator as well.

\paragraph{Chosen noise simulation approach}
\label{sec:noise_sim}
In this work, we use a simplified version of the noise model used in XCIST \cite{wu2022xcist}:
\begin{align}
    I_{i} &=  f_{CONV}\sum_k E_k\cdot\mathcal{P}(DQE_{ik}\cdot(A_{ik}+S_{ik}))+\mathcal{N}(\sigma_{electronic})\\
    I &= \Gamma_{\sigma_{cross-talk}}[I_1, I_2, ..., I_I]^T
\end{align}

where $i$ is the pixel index of the detector $I$, $E_k$ is the energy level with energy index $k$, $A_{ik}$ are the incident photons in the pixel, $S_{ik}$ the scattered photons in the detector. $DQE_{ik}$ is the detector quantum efficiency and $f_{CONV}$ the energy to electron conversion rate. The noise process is described by $\mathcal{P}$, a Poisson random generator, and $\mathcal{N}(\sigma_{electronic})$ is a zero mean Gaussian random generator with standard deviation $\sigma_{electronic}$. Finally, $\Gamma_{\sigma_{cross-talk}}$ is a $\mathbb{R}^{D\times D}$ matrix that models detector cross-talk, defined as a fraction of the signal $\sigma_{cross-talk}$ that is shared between adjacent pixels. 
\\[6pt]
This describes a full model of the detector behaviour given incident photons. In XCIST, the incident photons can be simulated by a Monte Carlo particle simulation based on a known source energy spectra and material decomposition of a sample. If the precise behavior of the energy-integrating detector is well understood for each energy level, the parameters $f_{CONV}$, $DQE_{ik}$, and $E_k$ related to the conversion of incident photons to measurements can be incorporated. However, for machine learning applications, the physics simulation would demand an unreasonably high computational time (several years for a sufficiently large dataset), necessitating simplifications of the model. In particular, the approximations done in this work assume that the measurement photons were produced by a monochromatic source ($k=1$) and that there are no scattered photons measured ($S_k=0$). Additionally, both the detector quantum efficiency of the pixels $DQE_k$ and the photon-to-electron conversion rate $f_{CONV}$ are assumed to be equal to one, as a detector specific calibration of these values is unknown and not easily obtainable without specialized lab equipment. This means all photons reaching the detector are assumed to be measured and no loss of signal is present. 
\\[6pt]
Thus, the chosen noise simulation approach to model the final measurement in the detector $I_{D}$ is:
\begin{align}\label{eq:noise}
     I_i &= \mathcal{P}(A_i)+ \mathcal{N}(\sigma_{electronic})\\
     I &=  \Gamma_{\sigma_{cross-talk}}[I_1, I_2, ..., I_I]^T.
\end{align}

\subsection*{Training data}
\label{sec:training_data}
For the development of a ML-based denoising algorithm the most important element is adequate high-quality training data. In a supervised training framework that means that there are pairs of input and target data. The algorithm is trained on these data pairs and learns a mapping from the input images to the target images. Zainulina et al. \cite{zainulina2022self} concluded that such supervised deep learning methods show the best performance, but the requirement of paired images may not always be easy to accomplish. For the case of image denoising, this means noisy CT sinograms/reconstructions as an input and noise-free or “clean” CT sinograms/reconstructions as a target data.
\\[6pt]
In this paper, we will use the term "experimental noisy data" in reference to raw low-dose CT measurement data acquired by a real-world experimental CT system. The term "simulated noisy data" will be used for artificially generated data for which artificial noise was applied to "clean" raw measurement data.
\\[6pt]
For experimental noisy data, the creation of corresponding image pairs requires a careful acquisition design to avoid that the algorithms would also learn a transformation or change of image content. The exact same CT slice needs to be scanned twice which makes it necessary to change the acquisition settings without infringing with the scanned object. The five main influencing acquisition parameters for the noise level within the CT images have been identified as source current ($I$), source voltage ($V$), exposure time ($t$), number of projections ($n_{proj}$), and number of averaged images ($n_{avim}$) \cite{rodriguez2020review}. Overall, the quantum noise in the reconstructed CT images is then inversely proportional to the square root of the number of detected photons. The aforementioned factors have their individual proportional influence on this number, which is given by: $\frac{ V^{1.3} }{\sqrt{I\times t \times n_{proj} \times n_{avim}}}$
\\[6pt]
Analysing this formula, we can determine the relationship between the acquisition parameters and the corresponding noise level in the reconstructed CT slices. Since the used tube voltage $V$ not only influences the noise level but also changes the energy of the used X-ray photons, a change of this factor was omitted. The number of averaged images $n_{avim}$ could not be decreased further than one and since the scanner was already operated close to the shortest possible exposure time $t$, changing that parameter was also not feasible. To avoid artifacts due to insufficient sampling of the object we did not decrease the number of projections $n_{proj}$. Therefore, the tube current $I$ was the only feasible option to change and both the noisy and the clean CT scans were acquired with the exact same parameters except for the tube current. For the clean data this was $\SI{1000}{\micro\ampere}$ whereas the noisy data had a 30 times smaller tube current of $\SI{33.3}{\micro\ampere}$.
\\[6pt]
For simulated noisy data creating corresponding image pairs is more straightforward. Given the "clean" data acquisition, a modification of the noise model in Eq.~\ref{eq:noise} can be used to simulate artificial noise into the clean image. Given the noise-free incident photons $A_i$ and that the outcome of the Poisson process can be described as an addition $\mathcal{P}(A_i)=A_i+P_i$, we can rewrite Eq.~\ref{eq:noise} for the acquisition of clean data as:
\begin{align}
    I_i^{clean}=A_i+P^{clean}+\mathcal{N}(\sigma_{electronic}),
\end{align}
\noindent
where the assumption of $P^{clean}=0$ can be made. This is not strictly true, but for a sufficiently large incident photon count $A_i$ it is approximately true. For the noisy acquisition thus the following holds:
\begin{align}
    I_i^{noisy}&=A_i+P_i^{noisy}+\mathcal{N}(\sigma_{electronic}).\\
    I_i^{noisy}&=I_i^{clean}+P_i^{noisy}\\
    I_{D}^{noisy}&=I_{D}^{clean} + \Gamma_{\sigma_{cross-talk}}[P_1^{noisy}, P_2^{noisy}, ..., P_I^{noisy}]^T.
\end{align}
\noindent
To appropriately simulate the low-dose measurements $I_{D}^{noisy}$, the noise distribution part of the total signal $P_i^{noisy}$ has to be produced, i.e. the Poisson component of the noise. Technically, $A_i$ would be a different number of photons for the clean and noisy images, as the noise mostly arises from the low photon count in our experiments and simulations. However, direct measurement of photon counts is not available and thus direct extraction of this noise from measured data is not possible. Therefore, the noise is parameterized by multiplying the flat-field corrected sinogram $ILI\in [0,1]$ (see paragraph "Pre-processing") by a parameters corresponding to the number of photons in vacuum, $I_0$, and generating Poisson statistics from its result as
\begin{align}
    P_i^{noisy} &= I_0\cdot ILI_i^{clean} - \mathcal{P}(I_0\cdot ILI_i^{clean}).
\end{align}
In this model, $I_0$ is the parameter to control the level of noise added to the clean data, a lower value representing noisier data. Based on the value in the XCIST software \cite{wu2022xcist}, a $\sigma_{cross-talk}$ of 5\% of the signal is added. 

\subsection*{Method development}
\label{sec:meth_dev}
The noise simulation and the algorithms for learned denoising in this work have been developed in LION (Learned Iterative Optimization Networks)\cite{LION}, an open-source toolbox for learned tomographic reconstruction implemented in Python. With a designated data loader for the 2DeteCT dataset and with CT experiments set up in a reproducible way it serves as an environment for a standardized comparison of the methods described below.
\\[6pt]
For the learned denoising algorithms we selected two common convolutional neural networks (CNNs) for image processing tasks that have been used for both natural images but also for computed tomography images in particular: The generic U-Net \cite{ronneberger2015u} and the mixed-scale dense neural network (MSD-Net) \cite{pelt2018mixed}. The U-Net, originally developed for the segmentation of biomedical images, has been adopted in many fields as a baseline for image reconstruction based on neural networks. The MSD-Net has proven to be particularly effective for computed tomography \cite{pelt2018improving,leuschner2021quantitative}. Its three main advantages are as follows: First, it has an advanced neural network architecture that uses dilated convolutions instead of traditional scaling operations to learn features at different scales. Second, it uses significantly fewer feature maps and trainable parameters which makes training it less computationally demanding and reduces the risk of over-fitting. Third, it has been applied to denoising large tomographic images and it has been proven that it can be easily applied to similar problems with minimal changes \cite{pelt2018mixed}.

\subsection*{Comparison metrics}
\label{sec:comp_metrics}
To evaluate the performance of the CNNs trained on either experimental or simulated noisy data, we consider two main comparison cases. In the first case, we test the performance of the algorithms in the setting that they have been trained on, i.e. settings in which they are supposed to work well. This means that if an algorithm is trained for denoising simulated noisy data this situation is used to score their overall performance. The same holds true for algorithms trained on experimental noisy data. For this we compare the output of our learned denoisers to the "clean" target data using the comparison metrics described below. In the second case, we want to compare the performance of the algorithms in settings for which they have not been trained for. This serves the purpose of checking their generalization to other tasks. It answers the question whether the algorithm generalizes to another noise model and its severity. In other words, whether the learned algorithms can also denoise input data without being trained on the specific noise of that data. This is particularly interesting for the case in which the learned denoisers are trained on simulated data and applied to experimental noisy data.
\\[6pt]
Using these comparison cases, we require comparison metrics with which we can evaluate the performance of the algorithms. Namely, how close the denoised images obtained from these algorithms are to the ground truth images. These metrics have to be able to measure two qualities: How well does the algorithm recover the structure of the imaged object from the noisy data? How well does it restore a good signal with respect to the overall noise in the reconstructed image?
\\[6pt]
Two commonly used metrics for these tasks are the structural similarity (SSIM) \cite{wang2004image} and the peak signal-to-noise ratio (PSNR) \cite{girod1992psychovisual}. The SSIM is a metric that indicates in a range from 0.0 to 1.0 how similar the compared image is to a ground truth, where 1.0 means they are identical. The PSNR is a metric that calculates the ratio between the highest attainable value of a signal and the strength of corrupting noise that impacts the fidelity of the image. Higher values in both metrics indicate a better algorithm performance. It is worth noting that these two commonly used quantitative metrics, may not be suitable for tomographic reconstruction or scalar fields \cite{breger2024study,breger2024study2}. In reconstruction tasks such as CT imaging in medicine, PSNR and SSIM do not necessarily reflect a task-dependent better image \cite{gourdeau2022proper,maruyama2023properties}. Therefore, it is suggested that evaluations consider such downstream tasks of the imaging rather than solely relying on traditional metrics. Additionally, the unbounded nature of CT images poses challenges for metrics like PSNR and SSIM, as the range of pixel values can vary. Different approaches to evaluating reconstruction performance, such as clipping or preserving the result range, can significantly impact reported performance. However, these metrics are still commonly used for a quantitative assessment of images and are used in this work as well since we are interested in measuring performance differences rather than rating the performance itself.
\\[6pt]
In our performance analysis we follow previous work by Zeng et al. \cite{zeng2015simple} who argued that image artifacts due to beam hardening and photon-starvation are particularly difficult to evaluate meaningfully with quantitative metrics in the sinogram domain and need a visual inspection in the reconstructed image domain. Therefore, we also include a qualitative, visualization-based evaluation between the results of denoised low-dose CT scans in the reconstructed image domain.

\subsection*{Computational experiments}
\label{sec:comp_exp}
\begin{figure}[!ht]
	\centering
	\includegraphics[width=0.95\columnwidth]{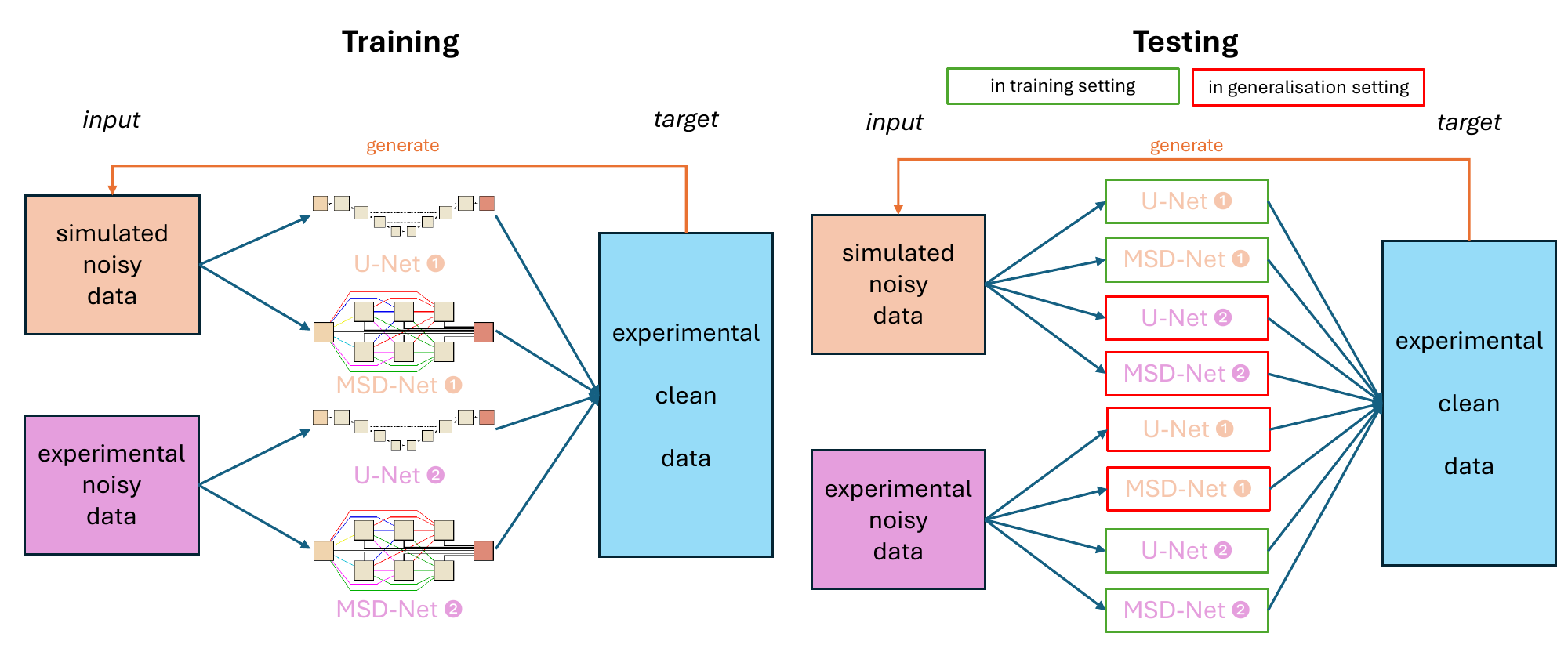}
    \caption{Training and testing scenarios for learned denoising networks (U-/MSD-Net illustrations adopted from\ \cite{pelt2018mixed}).}
	\label{fig:training_testing}
\end{figure}

For this work, we first applied the denoising in the projection domain, i.e. denoising beam intensity loss images ($ILI$), for three reasons: i) quality of denoising reconstructed images depends on the used reconstruction method; ii) artifacts caused by the noise in the projection domain are harder to remove after reconstruction; iii) noise in the projection domain is spatially uncorrelated. After evaluating the results of this approach we additionally trained denoising algorithms in an end-to-end setup, i.e., from sinogram to reconstruction. For this we included a FBP reconstruction in the pipeline of the models described below and visualized in Fig. \ref{fig:training_testing}.
\\[6pt]
We prepared the training data for our learned denoisers by generating simulated noisy data from the "clean" experimental measurement data as described in Section "Training data" and using the unchanged clean data as ground truth target data. Consequently, there are two respective image pairs for supervised learning available: First, the simulated noisy data as an input and the experimental clean data as a target. Second, the experimental noisy data as an input and the experimental clean data as a target. 
\\[6pt]
These image pairs were split into $\sim 80\%$ training data (3930 slices), $\sim 10\%$ validation (550 slices) and $\sim 10\%$ testing data (470 slices). Each algorithm was trained for 100 epochs using the Adam optimization algorithm \cite{kingma2014adam}. The final model parameters were selected based on minimal validation loss. The computations were carried out on a GPU-server with 4x RTX 2080Ti (11GB), 384GB RAM, and 2x 16-core Xeon CPUs as well as a GPU-server with 2x RTX A6000 (48GB), 1TB RAM, and 2x 16-core Xeon CPUs. 
\\[6pt]
After the training of the two neural network architectures on the two supervised learning settings, each of the four resulting trained networks was applied to their own test sets but also to the test sets of the data type they have not been trained on. A visual overview of this is given in Fig.~\ref{fig:training_testing}.

\section*{Results and Discussion}
\label{sec:results_disc}

\subsection*{Empirical Selection of Noise Level}
\begin{figure}[!ht]
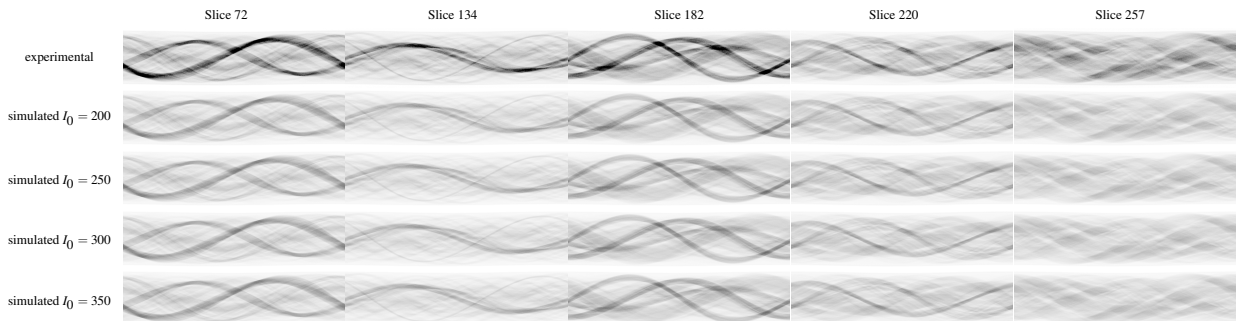

	\centering
    \begin{tikzonimage}[width=0.9\linewidth]{{"./Figure2"}.pdf}
    \node at (0.12, 1.015) {\tiny{Slice 72}};
    \node at (0.31, 1.015) {\tiny{Slice 134}};
    \node at (0.5, 1.015) {\tiny{Slice 182}};
    \node at (0.68, 1.015) {\tiny{Slice 220}};
    \node at (0.87, 1.015) {\tiny{Slice 257}};
    
    \node at (-0.02, 0.880) {\tiny{experimental}};
    \node at (-0.02, 0.690) {\tiny{simulated $I_0 = 200$}};
    \node at (-0.02, 0.490) {\tiny{simulated $I_0 = 250$}};
    \node at (-0.02, 0.300) {\tiny{simulated $I_0 = 300$}};
    \node at (-0.02, 0.110) {\tiny{simulated $I_0 = 350$}};
    \end{tikzonimage}
	\caption{Visual comparison of the sinograms of experimental and simulated noisy data with different levels of $I_0$ (200, 250, 300, 350) from the 2DeteCT dataset for the slices with indices 72, 134, 182, 220, 257.}
	\label{fig:simulation_sinoDomain}
\end{figure}

\begin{figure}[!ht]
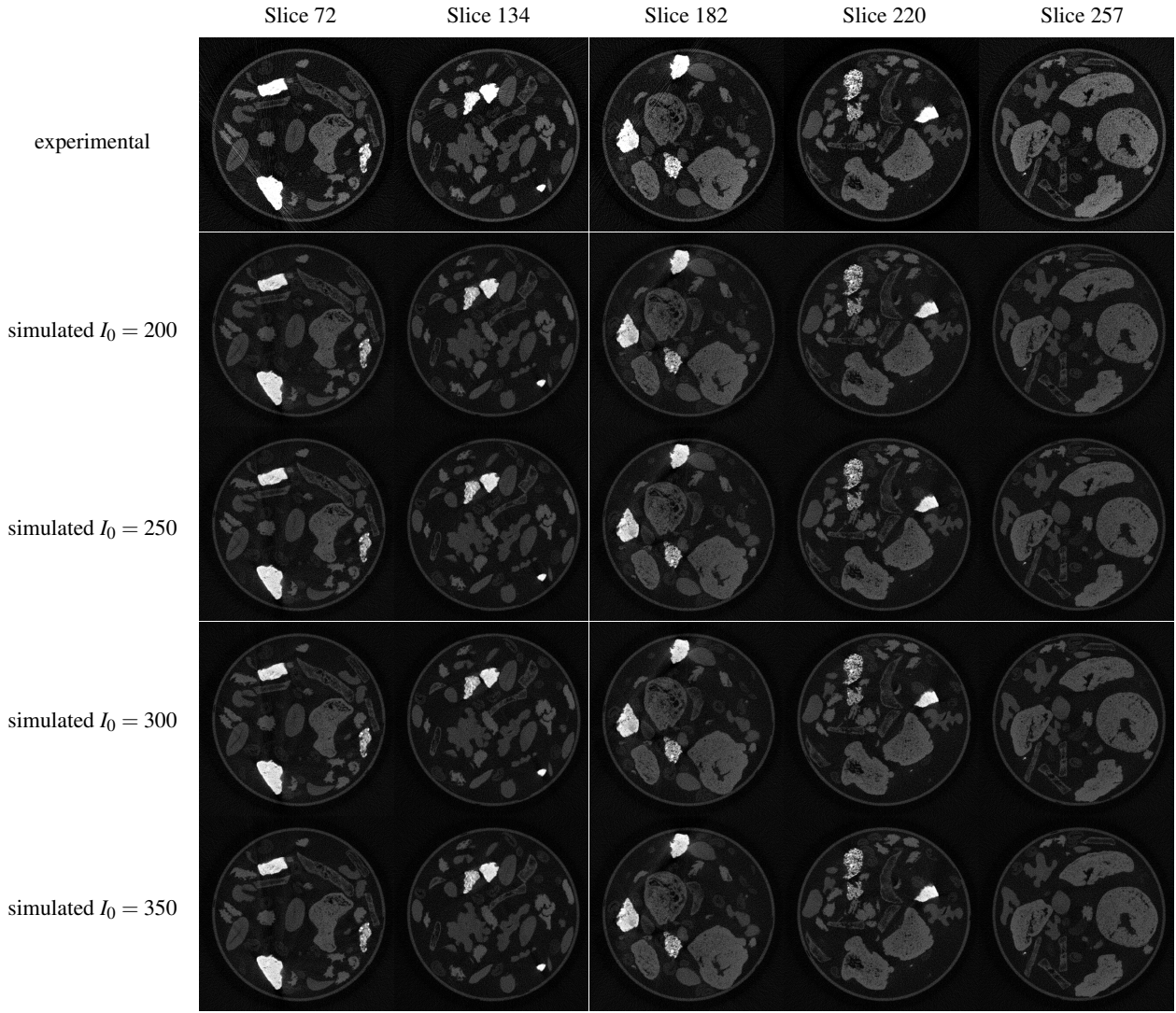

	\centering
    \begin{tikzonimage}[width=0.8\linewidth]{{"./Figure3"}.pdf}
    \node at (0.11, 1.015) {\small{Slice 72}};
    \node at (0.30, 1.015) {\small{Slice 134}};
    \node at (0.5, 1.015) {\small{Slice 182}};
    \node at (0.70, 1.015) {\small{Slice 220}};
    \node at (0.90, 1.015) {\small{Slice 257}};
    
    \node at (-0.1, 0.885) {\small{experimental}};
    \node at (-0.1, 0.695) {\small{simulated $I_0 = 200$}};
    \node at (-0.1, 0.495) {\small{simulated $I_0 = 250$}};
    \node at (-0.1, 0.300) {\small{simulated $I_0 = 300$}};
    \node at (-0.1, 0.110) {\small{simulated $I_0 = 350$}};
    \end{tikzonimage}
	
	\caption{Visual comparison of the FBP-reconstructed images of the experimental and simulated noisy data with different levels of $I_0$ (200, 250, 300, 350) from the 2DeteCT dataset for the slices with indices 72, 134, 182, 220, 257.}
	\label{fig:simulation_reconDomain}
\end{figure}

For our comprehensive study on the differences between the observed performances of algorithms trained on simulated noisy data and on experimental noisy data it was particularly important to have noise levels in our simulated noisy data that are representative of the noise levels present in our experimental noisy data. Therefore, we tried out various values of $I_0$ for our noise simulation approach and compared both the resulting simulated noisy data and the experimental noisy data to the "clean" sinogram data with respect to PSNR and SSIM. Furthermore, the quantitative comparison was also carried out in the reconstruction domain, i.e. comparing the FBP-reconstructed images of the experimental and simulated noisy data to the "clean" reference reconstructions of the 2DeteCT dataset. Observing similar numerical values w.r.t. PSNR and SSIM for both experimental and simulated noisy data we can argue that our noise model generates simulated noisy data with a similar noise level. The results of this comparison can be found in Table~\ref{table:empirical_noise_level}. They show that the SSIM and PSNR values in the sinogram domain are significantly larger than the respective experimental noisy data for all noise levels of the simulated noisy data. A visual comparison of the different noise levels in the sinogram domain proofed uninformative as displayed in Fig.~\ref{fig:simulation_sinoDomain}.
\\[6pt]
Corresponding quantitative and qualitative analyses in the reconstruction domain showed similar image metrics for both the simulated and experimental noisy data. For a noise level of $I_0 = 200$ the PSNR value is closest to the same metric for the experimental noisy data, whereas the SSIM value shows its best agreement for a noise level of $I_0 = 300$. Since the task at hand is learned denoising, we chose to rely on the agreement with respect to the PSNR value and chose a noise level of $I_0 = 200$ for our computational experiments. A qualitative inspection of the images in Fig.~\ref{fig:simulation_reconDomain} agrees with this choice of parameter.
\\[6pt]
A detailed inspection of Fig.~\ref{fig:simulation_reconDomain} furthermore showed a strong influence of the attenuation of the objects in each scan on the similarity between reconstructions based on simulated and experimental data. Simulated noisy image slices with no or only small objects with high attenuation (stones) appear to be visually close to the experimental noisy images. However, if those objects are bigger or grouped closely, the experimental noisy images show streaking artifacts caused by beam hardening, not visible in the reconstructions of the simulated noisy data. As previously mentioned, the noise model used in this work assumes mono-energetic photons and consequently cannot capture this behaviour. For the high-dose measurement data, which is the basis for the simulated noisy data, a high enough number of photons is detected and the reconstructed images do not present streaking artifacts due to beam hardening and some level of photon starvation. 

\begin{table*}
  \caption{Empirical selection of the appropriate noise level $I_0$ to generate the simulated noisy training data based on the SSIM and PSNR values of the data with respect to the ground truth (wrt GT) data of "mode 2".}
  \label{table:empirical_noise_level}
  \centering
  \resizebox{0.85\columnwidth}{!}{
  \begin{tabular}{lc cc cc}
    \toprule
    \multirow{2}{*}{\textbf{Type of Noisy Data}} & \multirow{2}{*}{\textbf{Noise Level}} & \multicolumn{2}{c}{\multirow{2}{*}{\textbf{Evaluation in Sinogram Domain}}} & \multicolumn{2}{l}{\textbf{Evaluation in Reconstruction Domain}}\\
    & & & & \multicolumn{2}{c}{\textbf{(FBP of Noisy Data)}}\\
    \midrule
    \midrule
    & & SSIM (wrt GT) & PSNR  (wrt GT) & SSIM  (wrt GT) & PSNR (wrt GT)  \\
    \midrule
    \textbf{Experimental Noise} & - & 0.2658 $\pm$ 0.0963 & 19.8130 $\pm$ 4.6583 & \textbf{0.1899 $\pm$ 0.0987} & \textbf{21.8024 $\pm$ 3.5996} \\
    \midrule
    \textbf{Simulated Noise} & $\mathbf{I_0 = 200}$ & 0.2965 $\pm$ 0.0409 & 25.7190 $\pm$ 0.8567 & 0.1364 $\pm$ 0.0307 & \textbf{21.4468 $\pm$ 1.8307} \\
    \midrule
    Simulated Noise & $I_0 = 250$ & 0.3448 $\pm$ 0.0435 & 26.6607 $\pm$ 0.8634 & 0.1627 $\pm$ 0.0356 & 22.3976 $\pm$ 1.8311 \\
    \midrule
    \textbf{Simulated Noise} & $\mathbf{I_0 = 300}$ & 0.3854 $\pm$ 0.0451 & 27.4191 $\pm$ 0.8678 & \textbf{0.1865 $\pm$ 0.0397} & 23.1659 $\pm$ 1.8317 \\
    \midrule
    Simulated Noise & $I_0 = 350$ & 0.4201 $\pm$ 0.0459 & 28.0517 $\pm$ 0.8725 & 0.2083 $\pm$ 0.0432 & 23.8089 $\pm$ 1.8325 \\
    \midrule
    \bottomrule
  \end{tabular}
  }
\end{table*}

\subsection*{Sinogram Denoising}
The quantitative analysis of the performance of the different CNNs trained on either experimental or simulated noisy data was carried out in both the sinogram and the reconstruction domain (FBP of model output) and is presented in Table~\ref{table:quantitative_analysis}. The evaluation in the sinogram domain shows that for both CNN architectures, U-Net and MSD-Net, the training on simulated noisy data performs better in both application cases, experimental and simulated noisy data. Applying the U-Net trained on experimental noisy data to simulated noisy data performs similarly well whereas the MSD-Net trained on experimental noisy data is not able to generalize well. Applying the U-Net trained on simulated noisy data to experimental noisy data yields both a 3dB lower PSNR and a 0.0522 lower SSIM. For the MSD-Net this gap is even more significant, with 21.7057 dB lower PSNR and 0.1323 lower SSIM. This might be due to the much lower number of parameters of the MSD-Net which is not able to capture the experimental noise equally well as the simulated artificial noise.
\\[6pt]
However, CT reconstruction is an inverse problem that can exacerbate noise from the sinogram during the reconstruction process. Furthermore, applying the required sinogram pre-processing steps changes the nature of the noise model in a complex way. Therefore, evaluating the performance of the denoisers in the reconstruction domain is scientifically more relevant and Table~\ref{table:quantitative_analysis} also compares the performance of the sinogram denoisers in the reconstruction domain (FBP of model output).
\\[6pt]
In there we can observe that the high performance in denoising the sinograms does not carry over to the reconstruction domain. Both the structural similarity and the PSNR in this domain drop substantially. Additionally, the evaluation in the reconstruction domain shows that learned denoising of experimental noisy data performs best if the CNNs are trained on experimental noisy data, as it is expected. Furthermore, the U-Net architecture seems to pick up the image content in terms of structural similarity (SSIM) better than the MSD-Net when trained on experimental noisy data. The PSNR performance is better for the MSD-Net in all training settings except for the case of training on simulated noisy data and testing on experimental noisy data.
\\[6pt]
After the uninformative visual inspection of the simulated noise in the sinogram domain, and considering that the ultimate goal is to obtain better reconstructed images, the qualitative analysis of the model performances was only carried out in the reconstruction domain which can be found in Fig.~\ref{fig:QualitativeAnalysis_reconDomain}. The qualitative visual inspection shows that the models for sinogram denoising (found within the first four rows of the figure) do not produce high-quality reconstructions, particularly regarding fine image features/details. The images exhibit lower noise than the FBP reconstructions of the noisy data directly, but there is a noticeable loss of image sharpness.

\begin{table*}
  \caption{Quantitative performance analysis with PSNR and SSIM of the differently trained models in the reconstruction domain for the two different testing data with respect to the ground truth data from the iterative reference reconstructions of "mode 2" from the 2DeteCT dataset.}
  \label{table:quantitative_analysis}
  \centering
  \resizebox{0.57\columnwidth}{!}{
  \begin{tabular}{lll cc}
    \toprule
    \multirow{2}{*}{\textbf{Method}} & \multirow{2}{*}{\textbf{Training Data}} & \multirow{2}{*}{\textbf{Metric}} & \multicolumn{2}{c}{\textbf{Testing Data}}                   \\
     & & & Experimental Noisy Data & Simulated Noisy Data \\
    \midrule
    \midrule
    \multicolumn{4}{l}{\textbf{Evaluation in Sinogram Domain}} \\
    \midrule
    \multirow{2}{*}{U-Net \circled{2}} & \multirow{2}{*}{Experimental Noisy Data} & SSIM 
    & 0.8126 $\pm$ 0.0194 & 0.8167 $\pm$ 0.0199 \\     
    & & PSNR & 18.4966 $\pm$ 0.6278 & 19.3181 $\pm$ 0.5575 \\
    \midrule
    \multirow{2}{*}{U-Net \circled{1}} & \multirow{2}{*}{Simulated Noisy Data} & SSIM 
    & 0.8273 $\pm$ 0.0240 & 0.8795 $\pm$ 0.0206 \\     
    & & PSNR & 33.4602 $\pm$ 0.9533 & 36.6016 $\pm$ 0.5616 \\
    \midrule
    \multirow{2}{*}{MSD-Net \circled{2}} & \multirow{2}{*}{Experimental Noisy Data} & SSIM
    & \textbf{0.8613 $\pm$ 0.0211} & 0.8239 $\pm$ 0.0216 \\     
    & & PSNR & \textbf{36.2182 $\pm$ 0.7214} & 20.4747 $\pm$ 0.6793 \\
    \midrule
    \multirow{2}{*}{MSD-Net \circled{1}} & \multirow{2}{*}{Simulated Noisy Data} & SSIM  
    & 0.7512 $\pm$ 0.0226 & \textbf{0.8835 $\pm$ 0.0198} \\     
    & & PSNR & 16.3208 $\pm$ 1.3965 & \textbf{38.0265 $\pm$ 0.7412} \\
    \midrule
    \midrule
    \multicolumn{4}{l}{\textbf{Evaluation in Reconstruction Domain (FBP of model output)}} \\
    \midrule
    \multirow{2}{*}{U-Net \circled{2}} & \multirow{2}{*}{Experimental Noisy Data} & SSIM 
    & \textbf{0.6134 $\pm$ 0.0732} & 0.6273 $\pm$ 0.0717 \\     
    & & PSNR & 26.7127 $\pm$ 1.9780 & 27.5290 $\pm$ 1.9405 \\
    \midrule
    \multirow{2}{*}{U-Net \circled{1}} & \multirow{2}{*}{Simulated Noisy Data} & SSIM
    & 0.5504 $\pm$ 0.0677 & 0.6351 $\pm$ 0.0713 \\     
    & & PSNR & 28.3307 $\pm$ 2.0810 & 32.5568 $\pm$ 2.0169 \\
    \midrule
    \multirow{2}{*}{MSD-Net \circled{2}} & \multirow{2}{*}{Experimental Noisy Data} & SSIM 
    & 0.5984 $\pm$ 0.0741 & 0.6152 $\pm$ 0.0723 \\     
    & & PSNR & \textbf{30.9185 $\pm$ 1.9707} & 28.3031 $\pm$ 1.9314 \\
    \midrule
    \multirow{2}{*}{MSD-Net \circled{1}} & \multirow{2}{*}{Simulated Noisy Data} & SSIM 
    & 0.3854 $\pm$ 0.0469 & \textbf{0.6372 $\pm$ 0.0469} \\     
    & & PSNR & 11.6366 $\pm$ 2.3636 & \textbf{32.6552 $\pm$ 2.0173} \\
    \midrule
    \midrule
    \multicolumn{4}{l}{\textbf{Evaluation in Reconstruction Domain (model output)}} \\
    \midrule
    \multirow{2}{*}{FBP+U-Net \circled{2}} & \multirow{2}{*}{Experimental Noisy Data} & SSIM
    & \textbf{0.8161 $\pm$ 0.0592} & 0.7466 $\pm$ 0.0681 \\     
    & & PSNR & 29.8398 $\pm$ 2.1834 & 28.0435 $\pm$ 2.0848 \\
    \midrule
    \multirow{2}{*}{FBP+U-Net \circled{1}} & \multirow{2}{*}{Simulated Noisy Data} & SSIM 
    & 0.5957 $\pm$ 0.0844 & 0.6693 $\pm$ 0.0820 \\     
    & & PSNR & 26.8841 $\pm$ 3.4800 & 28.8134 $\pm$ 3.9418 \\
    \midrule
    \multirow{2}{*}{FBP+MSD-Net \circled{2}} & \multirow{2}{*}{Experimental Noisy Data} & SSIM  
    & 0.7829 $\pm$ 0.0749 & 0.7892 $\pm$ 0.0731 \\     
    & & PSNR & \textbf{32.0684 $\pm$ 1.9309} & 32.0704 $\pm$ 1.9580 \\
    \midrule
    \multirow{2}{*}{FBP+MSD-Net \circled{1}} & \multirow{2}{*}{Simulated Noisy Data} & SSIM 
    & 0.7615 $\pm$ 0.0702 & \textbf{0.8204 $\pm$ 0.0567} \\     
    & & PSNR & 30.6211 $\pm$ 2.0626 & \textbf{33.1053 $\pm$ 2.0120} \\
    \bottomrule
  \end{tabular}
  }
\end{table*}

\subsection*{End-to-End Training from Sinogram to Reconstruction}
\begin{figure}[!ht]
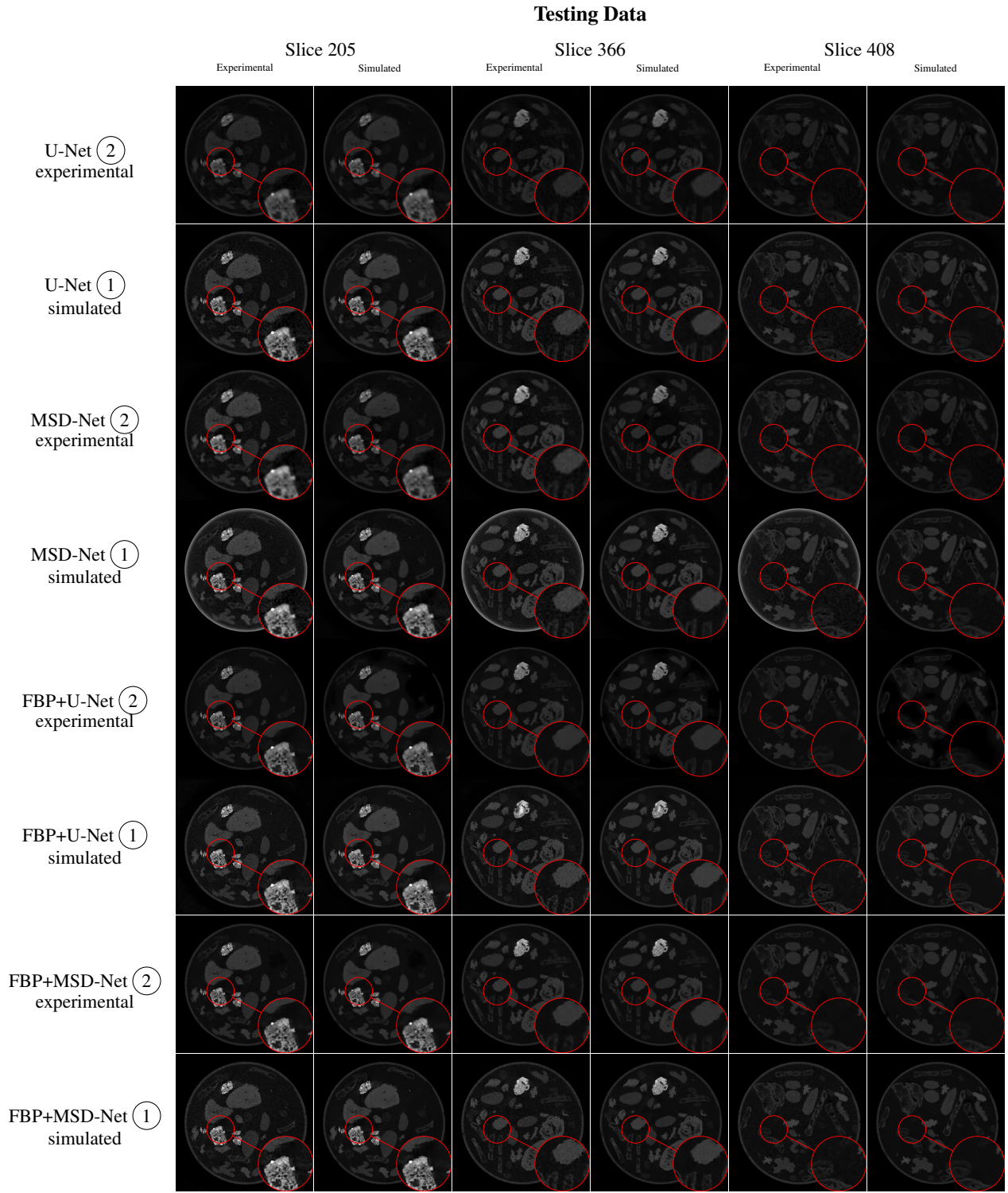

	\centering
    \begin{tikzonimage}[width=0.8\linewidth]{{"./Figure4"}.pdf}
    \node at (0.5, 1.055) {{\textbf{Testing Data}}};
    
    \node at (0.18, 1.025) {\small{Slice 205}};
    \node at (0.18-0.09, 1.008) {\tiny{Experimental}};
    \node at (0.18+0.07, 1.008) {\tiny{Simulated}};
    
    \node at (0.50, 1.025) {\small{Slice 366}};
    \node at (0.50-0.09, 1.008) {\tiny{Experimental}};
    \node at (0.50+0.08, 1.008) {\tiny{Simulated}};
    
    \node at (0.82, 1.025) {\small{Slice 408}};
    \node at (0.82-0.08, 1.008) {\tiny{Experimental}};
    \node at (0.82+0.09, 1.008) {\tiny{Simulated}};
    
    \node at (-0.1, 0.935) {\small{U-Net \circled{2}}};
    \node at (-0.1, 0.915) {\small{experimental}};
    \node at (-0.1, 0.815) {\small{U-Net \circled{1}}};
    \node at (-0.1, 0.795) {\small{simulated}};

    \node at (-0.1, 0.695) {\small{MSD-Net \circled{2}}};
    \node at (-0.1, 0.675) {\small{experimental}};
    \node at (-0.1, 0.575) {\small{MSD-Net \circled{1}}};
    \node at (-0.1, 0.555) {\small{simulated}};

    \node at (-0.1, 0.445) {\small{FBP+U-Net \circled{2}}};
    \node at (-0.1, 0.425) {\small{experimental}};
    \node at (-0.1, 0.325) {\small{FBP+U-Net \circled{1}}};
    \node at (-0.1, 0.305) {\small{simulated}};

    \node at (-0.1, 0.195) {\small{FBP+MSD-Net \circled{2}}};
    \node at (-0.1, 0.175) {\small{experimental}};
    \node at (-0.1, 0.075) {\small{FBP+MSD-Net \circled{1}}};
    \node at (-0.1, 0.055) {\small{simulated}};
    
    \end{tikzonimage}
	\caption{Qualitative performance analysis of the differently trained models in the reconstruction domain for the two different testing data  (slices indices 205, 366, 408).}
	\label{fig:QualitativeAnalysis_reconDomain}
\end{figure}

Having observed that a good model performance in the sinogram domain does not necessarily carry over to the reconstruction domain we wondered whether training the denoising algorithms in an "end-to-end" fashion, i.e. from noisy sinograms to "clean" reconstructions, would prove more effective as well. The model performance w.r.t. clean target reconstructions can be found in the bottom third of Table~\ref{table:quantitative_analysis} and in the bottom half of Fig.~\ref{fig:QualitativeAnalysis_reconDomain}. The relative performance of the networks for the respective combinations of training and testing settings is the same as before, but the results are substantially better. We observe an increase of 0.2027 in the SSIM for the best performing model in the constellation experimental noisy training data and experimental noisy testing data and an increase of 0.1832 in the SSIM for the best performing model in the constellation simulated noisy training data and simulated noisy testing data. Also the performance with respect to the PSNR for each corresponding constellation of training and testing data is better if the models are trained in an end-to-end fashion.
\\[6pt]
A qualitative analysis of the images in Fig.~\ref{fig:QualitativeAnalysis_reconDomain} show that the performance drop of training on simulated noisy data but testing on experimental noisy data is more substantial than what the performance metrics would suggest, as these metrics capture global performance rather than local. In all of the slices inspected, this particular train/test case produces the worst images of the quadruplet, for both models. Increased "graininess" permeates the entire image, and the low-intensity objects appear more porous than expected.

\begin{figure}[!ht]
	\centering
    \begin{tikzonimage}[width=0.8\linewidth]{{"./Figure5"}.pdf}
    \node at (0.5, 1.355) {{\textbf{Reference Reconstructions}}};
    
    \node at (0.20, 1.175) {\small{Slice 205}};
    \node at (0.20-0.07, 1.038) {\tiny{Experimental FBP}};
    \node at (0.20+0.07, 1.038) {\tiny{Ground Truth}};
    
    \node at (0.50, 1.175) {\small{Slice 366}};
    \node at (0.50-0.07, 1.038) {\tiny{Experimental FBP}};
    \node at (0.50+0.07, 1.038) {\tiny{Ground Truth}};
    
    \node at (0.8, 1.175) {\small{Slice 408}};
    \node at (0.8-0.07, 1.038) {\tiny{Experimental FBP}};
    \node at (0.8+0.07, 1.038) {\tiny{Ground Truth}};
    
    \end{tikzonimage}
	\caption{Reference reconstructions for the qualitative performance analysis of the differently trained models in the reconstruction domain for the two different testing data  (slices indices 205, 366, 408).}
	\label{fig:QualitativeAnalysis_Reference_reconDomain}
\end{figure}

\section*{Discussion and Conclusions}
\label{sec:conclusion}
In this work, we aimed to answer the question to which extent algorithms trained on simulated noisy data are applicable to real-world experimental noisy data. This was achieved through the implementation of a realistic yet computationally efficient simulation method and utilizing less-commonly available raw experimental measurement data.
\\[6pt]
After tuning the noise simulation to the experimentally measured noise level, our empirical selection of $I_0$ to set the noise level proved to be an adequate choice both in the qualitative and quantitative assessment (PSNR and SSIM) in the reconstruction domain. Differences in the simulation were mainly observed in the presence of large or closely grouped high-attenuation samples in the respective image slices, i.e. when beam hardening is present. This is expected, as the chosen noise model for simulation assumes monochromatic sources and thus cannot simulate highly non-linear effects such as beam hardening. 
\\[6pt]
While sinogram denoising achieved better results with simulated noisy data when evaluated in the sinogram domain, the performance did not carry over to the reconstruction domain where training on experimental noisy data showed a higher performance in denoising experimental noisy data. As previously mentioned, this is caused by the inherent ill-posedness of CT reconstruction, that amplifies any remaining noise in the process. Therefore, training the denoising algorithms in an end-to-end fashion from sinogram to reconstruction showed significant improvements in model performance, especially in terms of structural similarity and qualitative visual inspections of the reconstructions. It seems that the artifacts introduced by the FBP reconstruction are not too severe to mitigate via the subsequent post-processing network.
\\[6pt]
Our findings highlight the importance of carefully designing a noise simulation approach and choosing appropriate noise levels that match experimental data well. If possible the training should be conducted in an end-to-end fashion, i.e. mapping from raw measurement data to desired target reconstructions. In machine learning for computational imaging, simulated data can be quite different from experimental data, which can impact the transfer of learned systems to the real-world. In particular, the distributions of the training and testing data should be as close as possible and therefore training on experimental noisy data, if available, is preferable when the models are subsequently applied to experimental data. In our experiments, models trained on simulated data exhibit a measurable quantitative performance drop from simulated noisy testing data to experimental noisy testing data. This is even more noticeable by qualitative visual inspection, because these models produce the noisiest images from all the cases. 
\\[6pt]
As discussed before, our simulation model already captures much of the complexity of the experimental noise in the measurements. However, this work shows that the non-linearity of the imaging process is not captured well enough and that future work should investigate computationally efficient ways of including effects such as beam hardening or photon starvation. Possibly, generative models trained on experimental noisy and "clean" data could solve this challenge or alternatively simplified Monte Carlo particle simulations could be investigated. This study can serve as a starting point for crafting and testing even more sophisticated noise simulation approaches that might be able to close the sim-to-real gap \cite{nikolenko2021synthetic,weiss2024simulation} for CT image denoising.
\\[6pt]
Ultimately, this research shows that appropriately simulating real noise is important in learned CT research. While computationally fast noise models, like the one presented in this work, will produce data that are close enough to experimental data to make the models transferable to real-world applications, a drop in performance is expected. Hence, it is advisable to utilize real-world experimental data for training learned denoisers whenever feasible, and exercise caution in presenting performance outcomes solely based on simple performance metrics when training only on simulated noisy data.

\section*{Code availability}
\label{sec:code_availability}
Python scripts for setting up the neural network training as well as the evaluation of the noise reduction performance in the way described above are published on GitHub: \hyperlink{https://github.com/CambridgeCIA/LIONscripts/paper_scripts/noise_paper}{https://github.com/CambridgeCIA/LIONscripts/paper\_scripts/noise\_paper}. They make use of the ASTRA toolbox \cite{PALENSTIJN2011250,VANAARLE201535,vanAarle16}, which is openly available on (\hyperlink{ www.astra-toolbox.com }{ www.astra-toolbox.com }) and tomosipo \cite{hendriksen2021tomosipo}.

\section*{Acknowledgements} 
We are grateful to TESCAN-XRE NV, for their collaboration regarding the FleX-ray Laboratory. This work was supported by the Dutch Research Council (NWO, project numbers OCENW.KLEIN.285, 613.009.106, 639.073.506). A.B. acknowledges the support of the Accelerate Programme for Scientific Discovery and EPSRC grant EP/W004445/1. C.B.S additionally acknowledges support from the Philip Leverhulme Prize, the Royal Society Wolfson Fellowship, the EPSRC advanced career fellowship EP/V029428/1, the EPSRC programme grant EP/V026259/1, and the EPSRC grants EP/S026045/1 and EP/T003553/1, EP/N014588/1, the Wellcome Innovator Awards 215733/Z/19/Z and 221633/Z/20/Z, the European Union Horizon 2020 research and innovation programme under the Marie Skodowska-Curie grant agreement No. 777826 NoMADS, the Cantab Capital Institute for the Mathematics of Information and the Alan Turing Institute. This research was supported by the NIHR Cambridge Biomedical Research Centre (NIHR203312). The sponsors were not involved in the research and writing process.

\section*{Author contributions statement}
M.B.K. and A.B. contributed equally to this work. M.B.K. wrote the original draft of the manuscript and conceptualized the study and designed the experiments together with A.B. and F.L.. M.B.K. and A.B. set up all the computational experiments and performed all computations. A.B., F.L., C.B.S., and K.J.B. reviewed and edited the manuscript. All authors read and approved the final manuscript. 

\section*{Competing interests} 
The authors declare no competing interests.

\bibliography{Noise}

\end{document}